\documentclass[conference]{IEEEtran}
\IEEEoverridecommandlockouts
\usepackage{cite}
\usepackage{amsmath,amssymb,amsfonts}
\usepackage{algorithmic}
\usepackage{graphicx}
\usepackage{textcomp}
\usepackage{xcolor}
\usepackage{slashbox}
\def\BibTeX{{\rm B\kern-.05em{\sc i\kern-.025em b}\kern-.08em
    T\kern-.1667em\lower.7ex\hbox{E}\kern-.125emX}}
\begin{document}

\title{Extending TCP for Accelerating Replication on Cluster File Systems over SDNs}

\author{\IEEEauthorblockN{Sungheon Lim}
\IEEEauthorblockA{\textit{Computer Science and Engineering} \\
\textit{Korea University}\\
Seoul, Republic of Korea \\
lsh2240@korea.ac.kr}
\and
\IEEEauthorblockN{Hyogon Kim}
\IEEEauthorblockA{\textit{Computer Science} \\
\textit{Korea University}\\
Seoul, Republic of Korea \\
hyogon@korea.ac.kr}
}

\maketitle

\begin{abstract}
This paper explores the changes required of TCP to efficiently support cluster file systems such as Hadoop Distributed File System (HDFS) where the storage nodes are connected through a software defined networking (SDN). Traditional chain replications in these file systems incur large delay and cause inefficient network use. But SDN can cooperate with the cluster file systems to address the problems by pre-arranging a distribution tree, which opens the possibility of parallel replication. Unfortunately, it cannot be realized without extending TCP, to accommodate the parallel transfer on the transport layer. This paper discusses how to extend TCP to make it possible, and demonstrates the feasibility by implementing a prototype in the Linux kernel. The prototype saves the data replication time by 25\% while substantially reducing network use.
\end{abstract}

\begin{IEEEkeywords}
TCP extension, Software Defined Networking (SDN), cluster file system, data replication, network utilization, HDFS
\end{IEEEkeywords}

\section{Introduction}
Today, the data volume stored in data centers is growing explosively. Managing large volumes of data inevitably requires highly automated and efficient technologies. Prominent technologies among them are the large-scale cluster file systems such as Hadoop Distributed File System (HDFS) \cite{HDFS}, Google File System (GFS) \cite{GFS}, and Windows Azure Storage \cite{azure}. Among the prime management concerns for the cluster file systems is data storing operation for its heavy demand. For instance, more than 50\% of Facebook tasks in data centers spend their time storing data \cite{facebook-storing-tasks} and the amount of traffic to store data accounts for 50\% of the total traffic produced in Hadoop clusters \cite{HDFSprob1} \cite{HDFSprob2}. In this paper, we show that there is an opportunity for optimizing the data storing operation by considering the cooperation of cluster file systems and the SDN, using HDFS as a guiding example for its wide adoption in many data-parallel infrastructures.

\begin{figure*}[ht]
\centering
\includegraphics[width=0.575\linewidth]{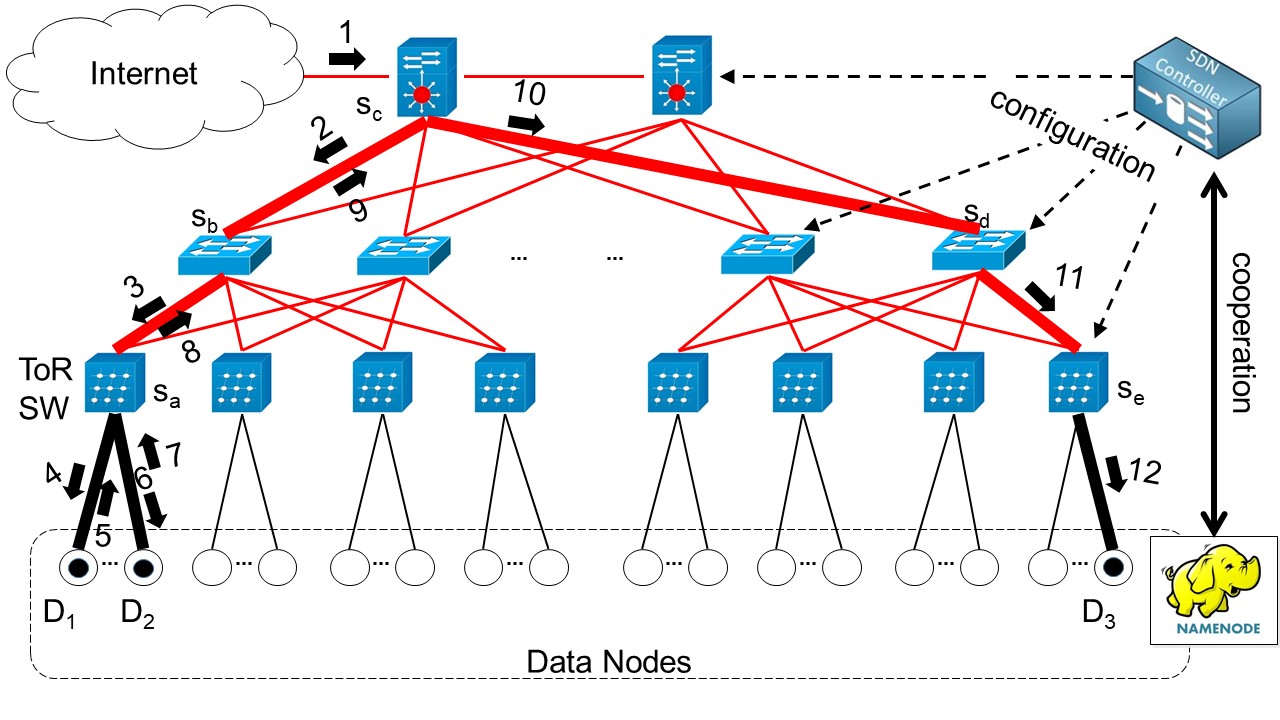}
\caption{Chain replication in HDFS: $D_1 \rightarrow D_2 \rightarrow D_3$ is the `pipeline'}
\label{fig:dc-topology}
\end{figure*}

In HDFS, fault tolerance is provided by replicating copies in multiple data nodes, with some copies situated on the same rack and others on different racks. Copies are made from node to node using a chain of TCP connections, or `pipeline' \cite{HDFS}. This traditional chain replication not only incurs sequential delay but also increases network traffic as the same data may traverse common network paths multiple times. In Figure \ref{fig:dc-topology}, for example, two of the copies are stored in the rack under the top-of-rack (ToR) switch $s_a$, and one copy in the rack under $s_e$. Suppose the client that writes the data is in the Internet. If the former copies are made first, the same data traverses twice the link between the data node $D_1$ and $s_a$ for the in-rack copies. Also, in the transfer from $D_2$ to $D_k$ for the third copy, three links (\textit{i.e.}, 7,8,9) are duplicately traversed. Even if other paths are taken, the number of traversed links remains the same.

If we let a SDN application consult with the Name Node of HDFS to identify the selected data nodes, however, we can open the possibility of parallel transfer, thereby reducing the cost incurred by the time-consuming and spatially inefficient serial transfers. Specifically, when the client asks the HDFS to store a file, the file system can in turn ask the SDN controller to create a distribution tree (\textit{e.g.} the thick edges in Figure \ref{fig:dc-topology}) among the involved SDN switches that connect the racks for the selected data nodes. Then the HDFS can transmit the data towards the first node in the pipeline for the chain replication as before, but the pre-arranged SDN distribution paths will mirror them to all data nodes simultaneously, saving time for itself and network resource for other tasks. This arrangement between the Hadoop Name Node and the SDN controller application can be made even in the current standard framework.


Sounding straightforward, however, such parallel transfer cannot be realized without extending TCP. This is because cluster file systems such as HDFS, GFS, and Windows Azure Storage use TCP as the transport protocol, and the SDN-level distribution is done on a lower layer than TCP. Thus the TCP receiver at $D_j (j \ne 1)$ would balk with a reset (RST) upon the data segment directly distributed from the client, as it establishes the connection not with the client but with $D_{j-1}$, its direct predecessor in the pipeline. Although one could imagine devising a specialized transport protocol \cite{HDFS_multicast} \cite{MCTCP} for this purpose, it would cause backward-compatibility problems. In other words, deployment base for today's widespread cluster file systems would need transition to the new protocol and the application programmers would need to adapt to the model provided by the new protocol. It would be prohibitively expensive, if not impossible. Therefore, we investigate how the TCP protocol can be extended to accommodate the parallel transfer model above, but no change required to the cluster file systems, not to mention the application programmers for them, using HDFS as a guiding example. Furthermore, we implement a proof-of-concept prototype in the Linux kernel, and demonstrate that it indeed leads to significant savings in both time and network use. Specifically, it improves the data transfer time by 25\% and saves 15 to 40\% traffic for the typical replication factor of 3. But first, we discuss below previous solutions for parallel transfer and their limitation.

\section{Related Work}
\subsection{Reliable Multicast Protocol}
There have been a bunch of researches on reliable multicast protocol for a long time, which provides a reliable sequence of packets to multiple recipients simultaneously. Previous reliable multicast solutions fall into three categories: network-based, sender-based, and receiver-based. ARM \cite{ARM} represents the first category (network-based), which requires the assistance from the network equipment such as routers to support reliable multicast. In ARM, routers suppress duplicate NACKs from multiple receivers to control the implosion problem in the upstream direction and limit the delivery of repair packets to receivers experiencing loss in the downstream direction. In the second category (sender-based), the sender maintains the receivers' information to initiate multicast (\textit{e.g.} TCP-XM \cite{TCP-XM} and MCTCP \cite{MCTCP}). Thus, it is usually suitable for the small group. TCP-SMO \cite{TCP-SMO}, RMTP \cite{RMTP} and RDCM \cite{RDCM} represent the last category (receiver-based), where the receivers need to acquire the multicast address in advance so that they can subscribe the multicast group. Especially, in RDCM \cite{RDCM}, the author emphasizes on reliable multicast for data centers and points out that existing reliable multicast solutions for the Internet are not suitable for the data center environment. It exploits diverse path existing in data center networks to build backup overlays for peer-to-peer packet repair.

\subsection{Multicast over SDNs}
Researches on multicast over SDNs usually focus on multicast routing and management to improve efficiency. MCDC \cite{MCDC} suggests multicast routing, in which a SDN controller selects the core routers with the least load. Avalanche \cite{Avalanche} proposes multicast system, leveraging SDN to take advantage of the rich path diversity commonly available in data center networks while minimizing the size of the routing tree for any given multicast group. Recently, more comprehensive multicast protocols over SDNs are proposed in MCTCP \cite{MCTCP} and ATHENA \cite{ATHENA}. The authors in MCTCP \cite{MCTCP} propose multicast TCP in SDN-based data centers. They exploit SDN as a means of managing IP multicast group and routing. The sender maintains each TCP connections with multiple receivers and moves its sliding window after the slowest receiver sends the acknowledgement (ACK). ATHENA \cite{ATHENA} enables reliable multicast and congestion control inside SDN based data centers so that no data is lost and other TCP flows in the network are not starved.

\subsection{Existing Problems in the Previous Solutions}

Previous solutions break the distributed, or `pipeline', structure originally designed for the cluster file systems and replace it with the multicast structure. 
To be specific, even the HDFS-specific protocols \cite{HDFS_multicast} \cite{MCTCP}, not to mention the reliable multicast protocols for the general purpose, insist the multicast structure not only for the data transfer, but also for the ACK transfer including the TCP ACK and the HDFS ACK, which is sent for every HDFS packet in the application layer. It causes realistic problems to exploit previous solutions for parallel transfer in today's widespread cluster file systems. Deployment base for them would need to transition to the new protocol and the application programmers would need to adapt to the model provided by the new protocol. It would be prohibitively expensive, if not impossible and cause backward-compatibility problems. Also, this structure makes the first sender, or the client, take over the distributed ACK processing responsibility from all data nodes in the pipeline, leading to the following problem.

From the perspective of the network, the cost for sending the ACKs becomes more expensive. Especially, the TCP incast problem at the gateway switch ($S_c$ in Figure \ref{fig:dc-topology}) would arise and TCP performance would degrade if the number of data nodes, trying to send the HDFS ACK to the client, reaches only six \cite{incast}, meaning two clients in the Internet, with the default replication factor of 3. This is because the HDFS ACK is the data from the perspective of TCP. Furthermore, the traveling paths for the ACK become longer, causing more delay. For example, $D_2$ in Figure \ref{fig:dc-topology} should change its destination for the ACK from $D_1$ to the client in the Internet.


\section{Background}
\subsection{Software Defined Networking (SDN)}
SDN is the architecture that decouples the control and data planes in the network. Control logic is centralized in the so-called SDN controller. By Exploiting this architecture, applications running on the SDN controller operate the network through flow-based forwarding decisions rather than destination-based ones \cite{BD_SDN}. OpenFlow \cite{openflow151}, or OpenFlow switch specification is the first standardized protocol in SDN and defines interfaces for the SDN controller to give commands to switches and to receive the network information from them. One of the most common behavior in this standard is the flow entry creation and matching process at SDN switches.

\begin{figure}[ht]
\centering
\includegraphics[width=0.9\linewidth]{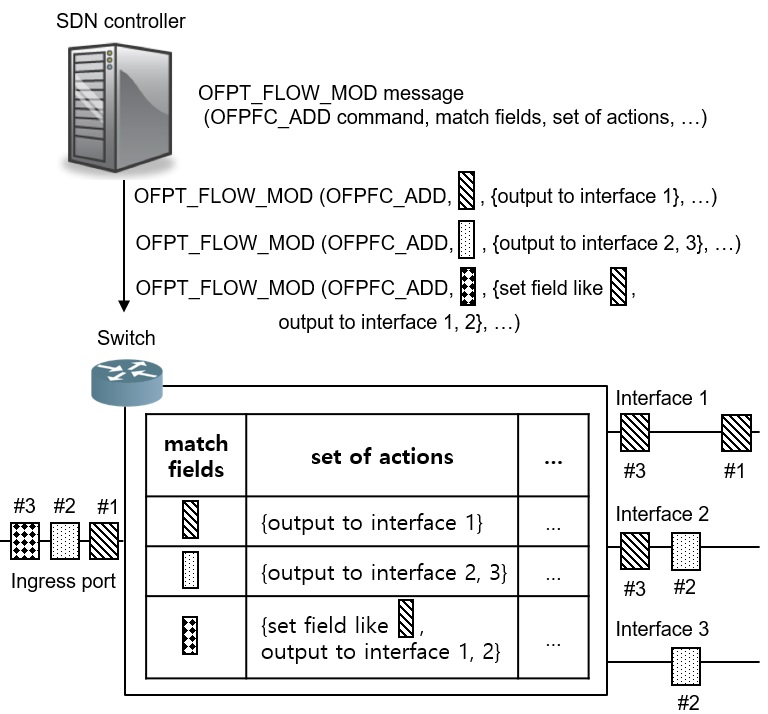}
\caption{Flow entry creation and matching process at the SDN switch \cite{openflow151}}
\label{fig:openflow}
\end{figure}

As depicted in Figure \ref{fig:openflow}, the SDN controller gives flow-based forwarding rules through \texttt{OFPT\_FLOW\_MOD} message, including \texttt{OFPFC\_ADD} command, matching field, and actions. Flow is identified by matching field represented as different rectangles in Figure \ref{fig:openflow}. Matching field contains IP address, ethernet address, TCP port numbers and so on. The switch does the action specified by the SDN controller to the identified flow. For example, it can modify TCP/IP header and forward the corresponding flow to the designated interface(s).

\subsection{Replication Process}
\begin{figure}[ht]
\centering
\includegraphics[width=0.85\linewidth]{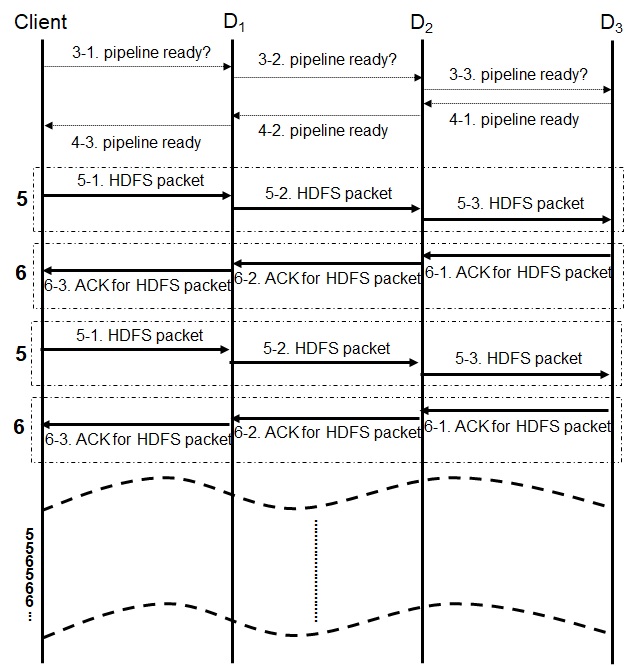}
\caption{Replication process in HDFS}
\label{fig:replication_HDFS}
\end{figure}

In order to provide the fault tolerance in the cluster file system, it stores the same data in multiple (\textit{i.e.,} 3) nodes. The cluster file system does this process in the application layer. Figure \ref{fig:replication_HDFS} shows the replication process in HDFS, which is used as a guiding example in this paper.

\begin{enumerate}
\item
Client requests the Name Node to store a block.
\item
The Name Node provides client with the data nodes to store the block.
\item
Client asks if $D_1$ is ready to receive the block (repeated until $D_3$).
\item
$D_3$ informs to $D_2$ that pipeline is ready (repeated until client).
\item
Client divides the block into multiple HDFS packets and send one of them to $D_1$ (repeated until $D_3$).
\item
$D_3$ informs to $D_2$ that it successfully received the HDFS packet (repeated until client).
\end{enumerate}

5 and 6 are repeated until the block transfer is completed. Even though this process may not exactly same in the other cluster file systems, this process is similar as long as the other cluster file systems use sequential TCP connections for replication like GFS and Windows Azure Storage.

\begin{figure*}[ht]
\centering
\includegraphics[width=0.7\linewidth]{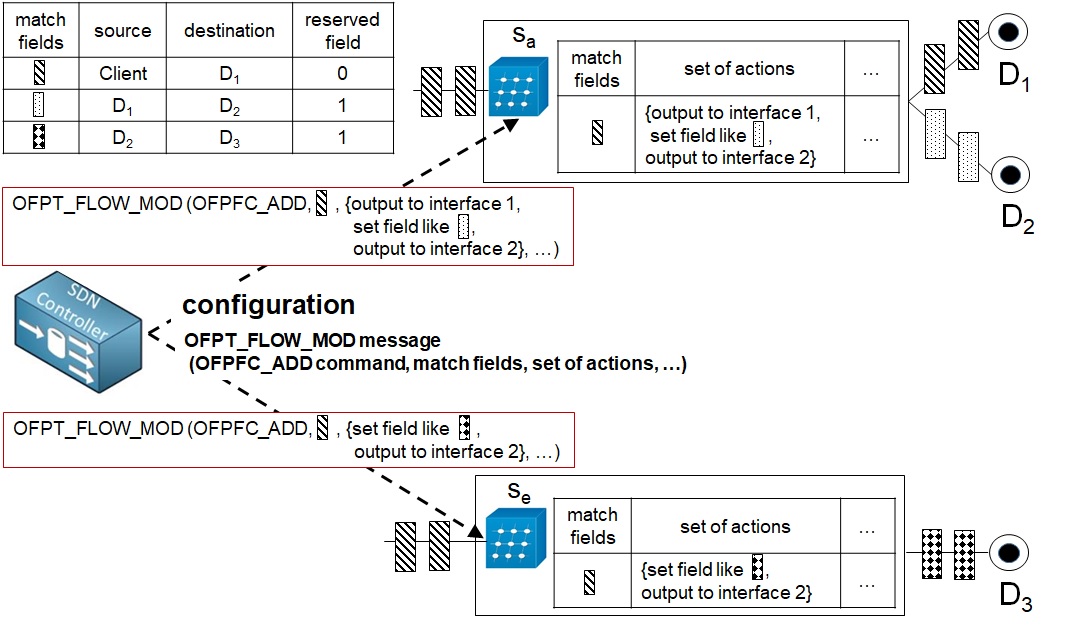}
\caption{Configuration for the transparent mirroring at SDN switches ($S_a$ and $S_e$) in Figure \ref{fig:dc-topology}}
\label{fig:configuration}
\end{figure*}

\section{Design}

\subsection{Challenges}
In order to solve the problems mentioned above, we extend TCP over SDNs, and the followings are the key issues in this paper.
\begin{itemize}
\item

Transparent mirroring with the SDN standard: Unlike previous solutions, our scheme focuses on solving the real-world problem, so it should be feasible in off-the-shelf SDN equipment, supporting the current standard framework.
\item
Successful reception of transparently mirrored TCP segments: The mirrored ones are not easy for the next data node in the pipeline to accept, because the TCP header information such as source, destination, and sequence number is not correct.

\item
Prevention of duplicate transmission of the transparently mirrored TCP segments: The previous data node in the pipeline tries to send the data already mirrored to the next data node, because the pipeline structure is intact. However, such an action is unnecessary in our scheme.

\item
Retransmission of the lost TCP segments: The previous data node in the pipeline does not need to send the data to the next data node, however, if the next data node lose the mirrored TCP segments, the previous data node are required to retransmit the TCP segments like in the original pipeline structure.

\end{itemize}

Next, we discuss below how the SDN controller application can configure SDN switches to make and forward the transparently mirrored TCP segments towards data nodes.

\subsection{Configuration for Mirroring at SDN Switches}
Suppose the Name Node selects the data nodes $\mathbf{D} = \{D_1, D_2, \ldots, D_k\}$ for storing a client-requested data, where $k$ is the replication factor. When the Name Node lets the SDN controller application know the pipeline information (IP addresses and TCP port numbers) for $\mathbf{D}$, it should create flow entries in the switches $\mathbf{S}$ connecting $\mathbf{D}$.
\subsubsection{Distributing the Mirrored TCP Segments}
At each switch in $\mathbf{S}$, it creates a flow entry that matches packets from the client to $D_1$, and performs \texttt{output} action(s) \cite{openflow151} on the interface(s) to those data nodes below the given switch. In order to compute the distributing output interface(s), the routing application running on the SDN controller is consulted to obtain the interfaces $I_c$ and $I_\mathbf{D}$ towards the client and $\mathbf{D}$, respectively. Then, forwarding interfaces are $I_\mathbf{D} - I_c$. For instance, $s_a$ in Figure \ref{fig:dc-topology} computes $\{I_{D_1},I_{D_2}\}$ as the forwarding interfaces. Table \ref{tbl:forwarding} summarizes forwarding interfaces at each switch in Figure \ref{fig:dc-topology}.

\begin{table}[ht]
\centering
\caption{Forwarding interfaces at each switches}
\label{tbl:forwarding}
\begin{tabular}{|c|c|c|c|}
\hline
\backslashbox{Switch}{Interface}   & $I_c$ & $I_\mathbf{D}$ & \begin{tabular}[c]{@{}c@{}}Forwarding\\ interface(s)\end{tabular} \\ \hline
$S_a$ & $I_{S_b}$  & $\{I_{D_1},I_{D_2}\}$  & $\{I_{D_1},I_{D_2}\}$                     \\ \hline
$S_b$ & $I_{S_c}$  & $\{I_{S_a}\}$  & $\{I_{S_a}\}$                     \\ \hline
$S_c$ & $I_{I}$  & $\{I_{S_b},I_{S_d}\}$  & $\{I_{S_b},I_{S_d}\}$                     \\ \hline
$S_d$ & $I_{S_c}$ & $\{I_{S_c},I_{S_e}\}$ & $\{I_{S_e}\}$                    \\ \hline
$S_e$ & $I_{S_d}$ & $\{I_{S_d},I_{D_3}\}$ & $\{I_{D_3}\}$                    \\ \hline
\end{tabular}
\end{table}

\subsubsection{Modifying TCP/IP Header for the Transparent Mirroring}
In addition to the \texttt{output} action, \texttt{set-field} actions \cite{openflow151} are configured in the ToR switches connecting mirroring data nodes $\mathbf{D}-\{D_1\}$ down the pipeline. For instance, if a switch needs to deliver the mirrored data to $D_j$, the \texttt{set-field} actions modify the IP addresses and the TCP port numbers of the client and $D_1$ in the source and the destination, respectively, to those of $D_{j-1}$ and $D_j$. Finally, one more \texttt{set-field} action is configured on the flow entry to set a flag in the reserved field of the TCP header for signaling the receiving TCP that the header has been changed and it is a mirrored copy.

For these configurations at SDN switches, the SDN controller application uses \texttt{OFPT\_FLOW\_MOD} message to carry the action-creating command \texttt{OFPFC\_ADD} \cite{openflow151}. Figure \ref{fig:configuration} shows configuration for the transparent mirroring at SDN switches ($S_a$ and $S_e$) in Figure \ref{fig:dc-topology}.


\subsection{TCP Extentions for Mirrored Replication}
\begin{figure}[ht]
\centering
\includegraphics[width=\linewidth]{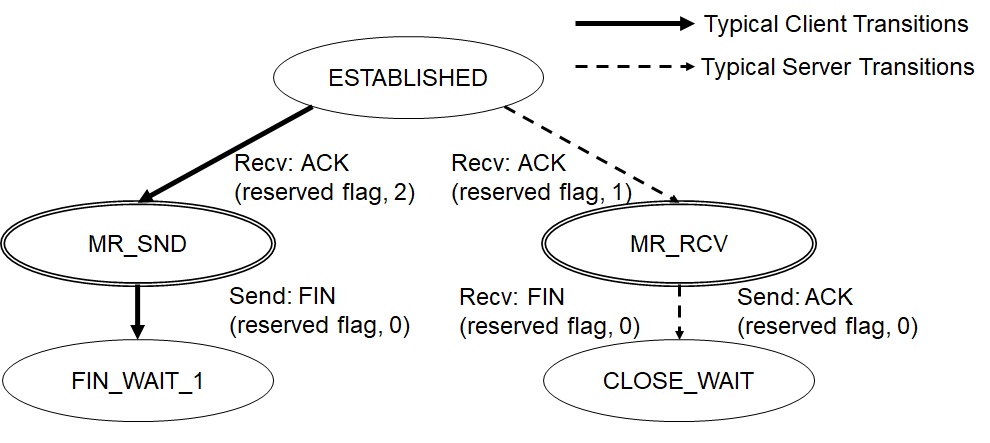}
\caption{The new states for TCP-MR in the TCP state transition diagram}
\label{fig:state}
\end{figure}

The central feature our modified TCP (henceforth TCP ``Mirrored Replication (MR)'') is that $D_j$ ($2 \le j \le k$) directly receives a HDFS data block from the client, not from the previous data node $D_{j-1}$ in the pipeline. To preserve the traditional HDFS chain replication abstraction, however, the pipeline should still be created and maintained. Thus in TCP-MR, the connection from $D_j$ is still established with $D_{j-1}$ and the TCP ACKs from $D_j$ is sent to $D_{j-1}$ (not to the client), as in the traditional HDFS pipeline. For this reason, when $D_j$ receives a TCP segment from the client (as if from $D_{j-1}$), it has to perform two operations in the newly added states (MR\_SND and MR\_RCV) in Figure \ref{fig:state}. First, it has to translate the sequence number and accept the payload in the received segment in the MR\_RCV state. Second, it should simulate transferring it to $D_{j+1}$ down the pipeline in the MR\_SND state, while not actually transmitting. Below, we discuss how these two operations are realized in TCP-MR.

\subsubsection{Receiving with Sequence Number Translation}
\begin{figure}[ht]
\centering
\includegraphics[width=0.7\linewidth]{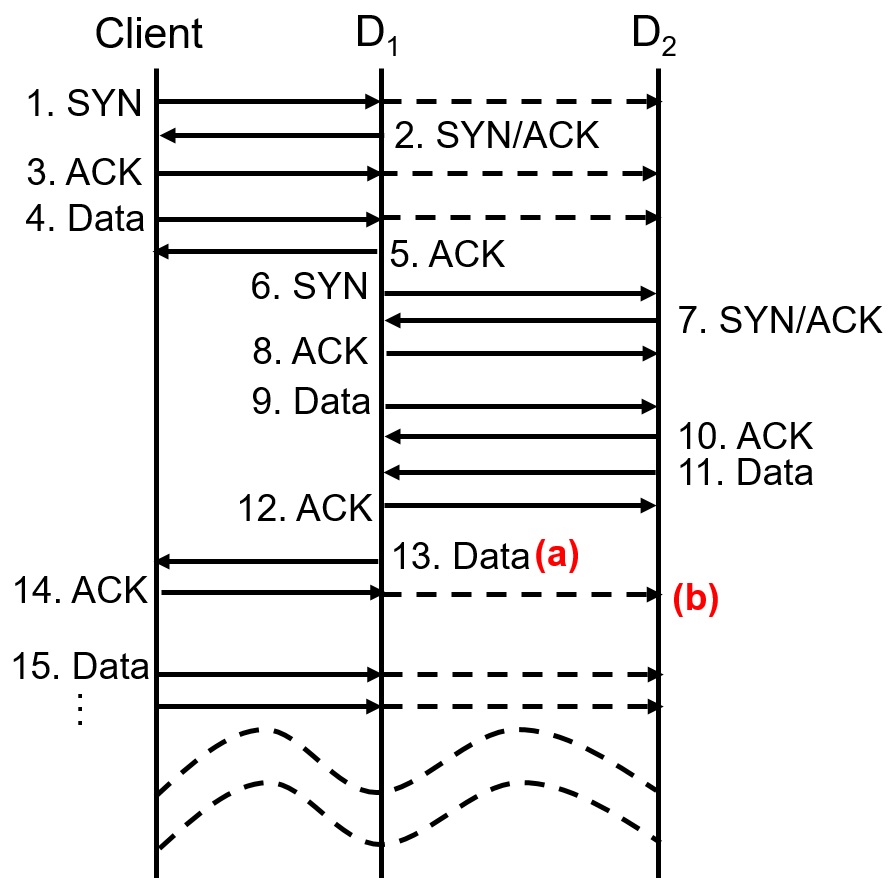}
\caption{The pipeline creation process \protect\cite{HDFS} from the perspective of TCP}
\label{fig:pipeline}
\end{figure}

\begin{figure}[ht]
\centering
\includegraphics[width=0.7\linewidth]{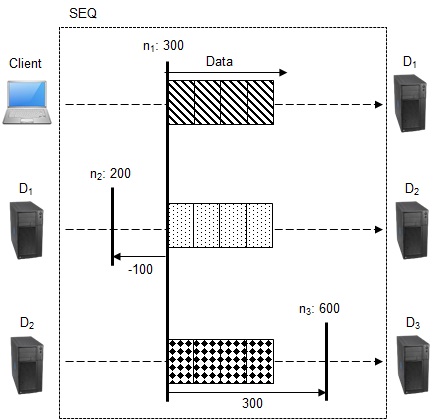}
\caption{An example of sequence number (SEQ) gap}
\label{fig:seq}
\end{figure}

\begin{figure}[ht]
\centering
\includegraphics[width=0.7\linewidth]{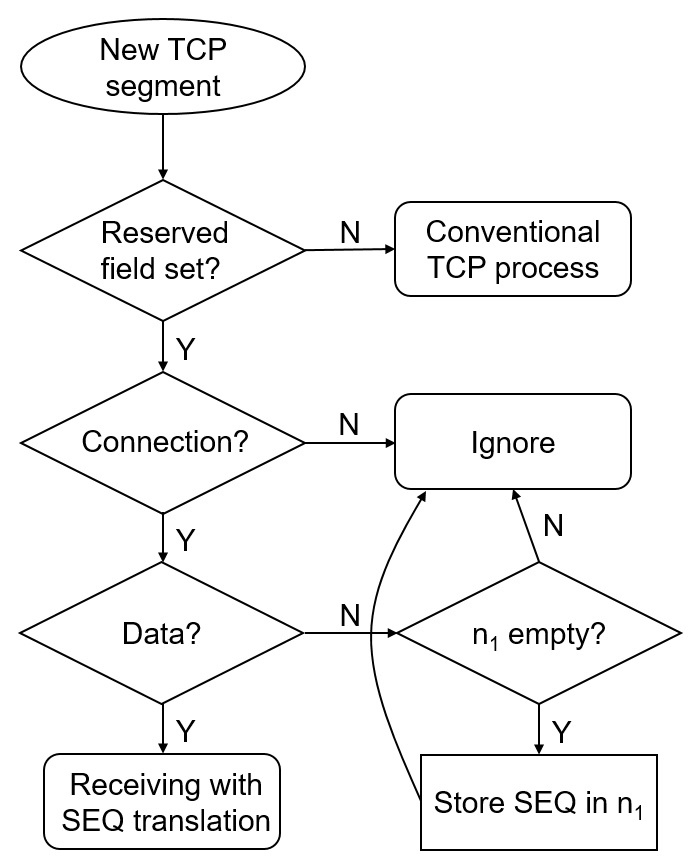}
\caption{The flow chart for the receiving process of $D_j$}
\label{fig:flowchart}
\end{figure}

The sequence number spaces are independent between the TCP connections in the pipeline. Figure \ref{fig:pipeline} shows the pipeline creation process \cite{HDFS} from the perspective of TCP (For $D_3$, similar procedure is repeated). At the last step of the pipeline completion, $D_1$ informs the client of it \cite{HDFS} (a in Figure \ref{fig:pipeline}), so that the client can start transmitting data (15 in Figure \ref{fig:pipeline}). But the client first acknowledges the notification-carrying TCP segment from $D_1$, which is mirrored to all $D_j$ ($2 \le j \le k$) (b in Figure \ref{fig:pipeline}). Then they use the ACK to compute the sequence number compensation given as
\begin{equation}\label{eq:seq}
\delta_j = n_j - n_1
\end{equation}
where $n_1$ is the sequence number in the mirrored ACK with the reserved flag set to 1,
and $n_j$ is the current sequence number of the TCP channel from $D_{j-1}$. For instance, $\delta_2$ is -100 and $\delta_3$ is 300 in Figure \ref{fig:seq}.
From this moment on, TCP-MR on $D_j$ enters into the MR\_RCV state in Figure \ref{fig:state} and accepts the data segments from the client that has the reserved flag set to 1, which signifies that the TCP data segment is a mirrored copy. Meanwhile, for other segments with the reserved flag set, the flags (SYN, ACK, FIN, RST, CWR, ECE) and the ACK number information in the TCP header are ignored because they are the signaling traffic between the client and $D_1$. But for the TCP segments from $D_{j-1}$, for example the retransmissions, $D_j$ processes them normally. Figure \ref{fig:flowchart} shows the flow chart for receiving process of $D_j$. Obviously, $D_j$ sends an ACK to $D_{j-1}$ as conventional TCP does except setting the reserved flag as 2. $D_{j-1}$ receiving the ACK with the reserved flag 2 enters into the MR\_SND state in Figure \ref{fig:state}.

\subsubsection{Simulating Forwarding Down the Pipeline}
In order to prevent $D_{j-1}$ ($2 \le j \le k$) from duplicating the client's transmission to $D_j$ and prepare necessary retransmission in the future, $D_{j-1}$ should slide the sending window, run the retransmission timer, and process the ACK from $D_j$, without actually transmitting data segments. We call this feature of TCP-MR the \textit{virtual transmission}, and it is done for non-retransmitted data segments. If the retransmission timer expires, however, $D_{j-1}$ should actually fill the hole for $D_j$ because the client should not engage with $D_j$ in the chain replication semantics. Table \ref{tbl:comparison} summarizes differences between TCP and TCP-MR from the perspective of $D_j$.

\begin{table}[ht]
\centering
\caption{Comparison between TCP and TCP-MR from the perspective of $D_j$ ($2 \le j \le k$)}
\label{tbl:comparison}
\begin{tabular}{|c|c|c|}
\hline
                                    & \textbf{TCP}    & \textbf{TCP-MR} \\ \hline
\begin{tabular}[c]{@{}c@{}}Non-retransmitted\\ data segment from\end{tabular} & $D_{j-1}$ & Client \\ \hline
\begin{tabular}[c]{@{}c@{}}Retransmitted\\ data segment from\end{tabular}     & \multicolumn{2}{c|}{$D_{j-1}$} \\ \hline
TCP ACK to                              & \multicolumn{2}{c|}{$D_{j-1}$} \\ \hline
\end{tabular}
\end{table}

Also note that since the data segment comes directly from the client, $D_{j-1}$ may receive an ACK from $D_j$ even before its virtual transmission in Figure \ref{fig:vtx}, if the following condition holds:

\begin{figure}[ht]
\centering
\includegraphics[width=0.7\linewidth]{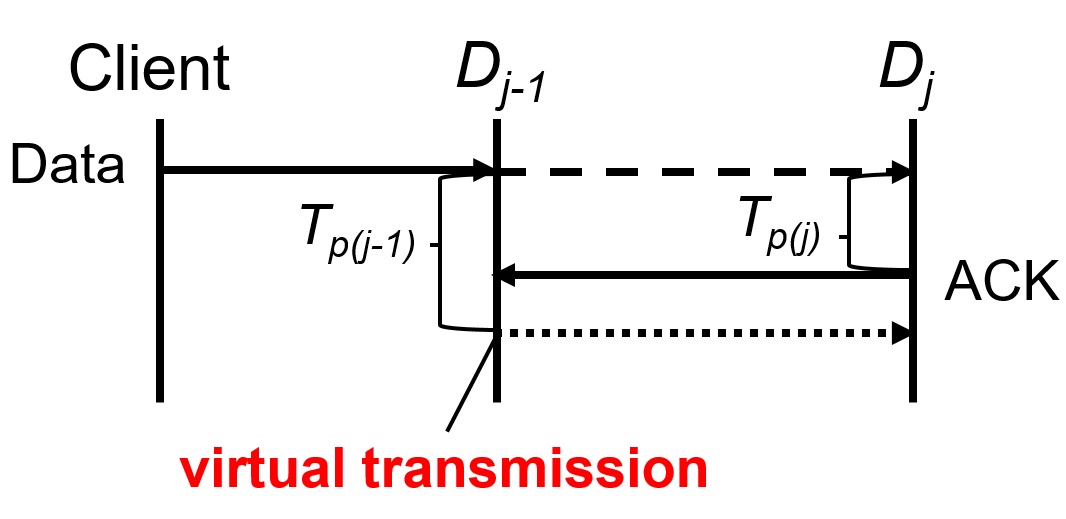}
\caption{Arrival of TCP ACK before virtual transmission}
\label{fig:vtx}
\end{figure}

\begin{equation}\label{eq:vtx}
T_{vtx} > T_{ack}
\end{equation}
\begin{equation}\label{eq:vtx1}
T_{vtx} = T_{c,j-1} + T_{p(j-1)}
\end{equation}
\begin{equation}\label{eq:vtx2}
T_{ack} = T_{c,j} + T_{p(j)} + T_{j,j-1}
\end{equation}

Here, $T_{c,j}$ is the time it takes for a TCP segment to travel from the client to $D_j$, and $T_{p(j)}$ is the time for $D_j$ to transmit an ACK after receiving the segment. $T_{p(j-1)}$ is the time for $D_{j-1}$ to perform the virtual transmission after receiving a TCP segment from the client. Note that $T_{p(j-1)} > T_{p(j)}$ can happen because $T_{p(j-1)}$ includes the time to receive multiple TCP segments to build a HDFS packet whose default size is 64 KB. In the original Hadoop model, $D_{j-1}$ starts to transfer a HDFS packet to $D_j$ only after it completely receives it and notifies the Hadoop application so that it may start the transfer \cite{HDFS}. In contrast, $T_{p(j)}$ includes only the reception and ACK generation delays, which could be much smaller than $T_{p(j)}$. In case $D_{j-1}$ receives an ACK from $D_j$ before it starts the virtual transmission, it first stores the ACK number and processes it upon the virtual transmission, as if it received a normal ACK with the number.



\section{Implementation and Performance Evaluation}
To prove the feasibility of the proposed concept and evaluate its performance, we implement it on a virtual machine platform using \texttt{VMware Workstation}. On a high-performance desktop with Intel \texttt{i7-6700K} CPU with 8 MB cache and 4.0 GHz clock, we create a virtual machine for each of the client, the data node, the SDN controller, and the Name Node. All are connected to a single SDN switch in a wheel-and-spoke topology. The SDN switch implements the \texttt{OpenFlow} switch in software as per \texttt{OpenvSwitch 2.5.0} \cite{ovs}, and \texttt{POX} \cite{pox} serves as the SDN controller. As for HDFS configurations, the HDFS block size is set to 128 MB, and HDFS packet size to 64 KB. The file size to be stored is set to 128 MB. For each data node, the Linux kernel (\texttt{Ubuntu 14.04}) is augmented with TCP-MR. We use two metrics for evaluation, HDFS block transfer time and network link use, as we vary the replication factor from 2 to 5. For performance evaluation, we set the maximum kernel memory size for data reception to a sufficient value to prevent throughput drop from having the window size at data nodes reduced to zero. Specifically, we set \texttt{writeMaxPackets} to 20 HDFS packets of size 64 KB each, so \texttt{net.core.rmem\_max} is set to 20$\times$64 KB = 1.3 MB.



\subsection{Block Transfer Time Reduction}
\begin{figure}[ht]
\centering
\includegraphics[width=0.95\linewidth]{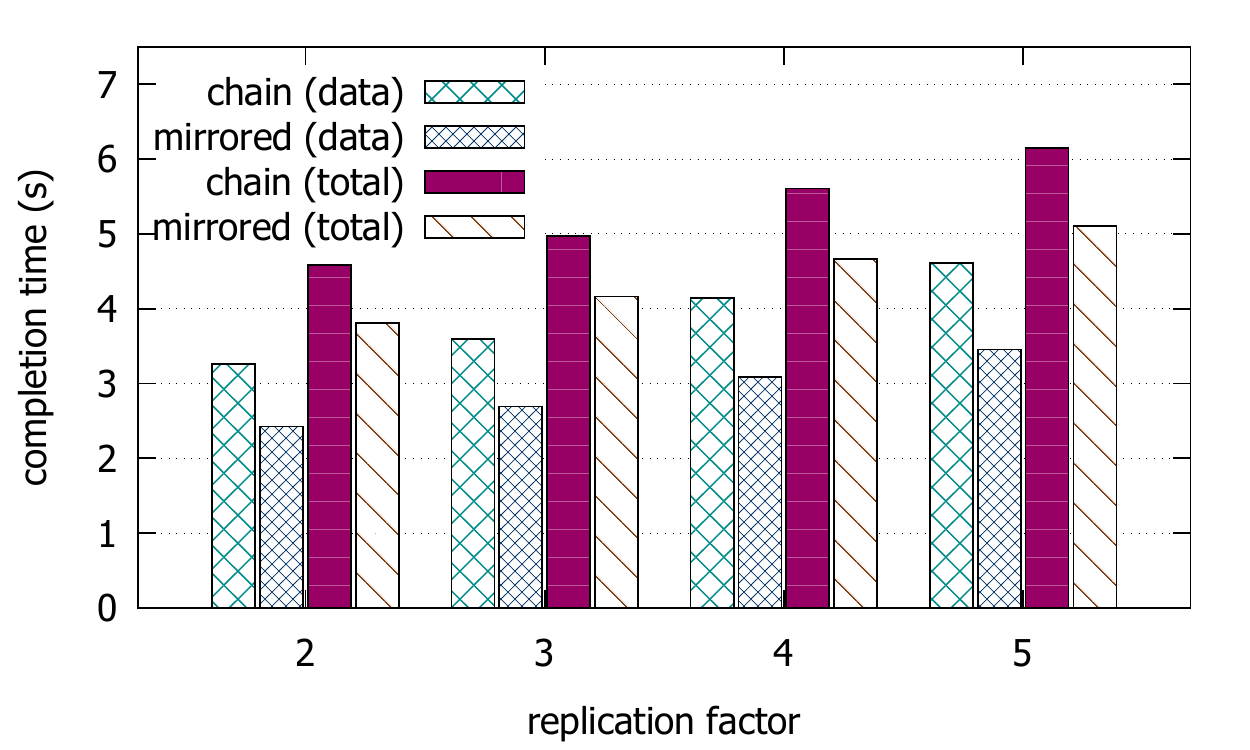}
\caption{Block transfer latency comparison as a function of the replication factor}
\label{fig:time}
\end{figure}

Figure \ref{fig:time} shows the transfer completion times for chain and mirrored replications. `Data' is the net data transfer time, and `total' is the total time including the pipeline setup, \textit{etc}. We observe that the implemented mirrored replication reduces the data transfer time by 25\%, and the total time by 17\%. Since the reduction applies to each HDFS data block, the ratio will be the same irrespective of the total file size.
Although the absolute measured times themselves on our simple virtualized testbed may not be too relevant because on a real data center network they can further depend on the network capacity and link congestion levels, the reduction ratios are a meaningful indication of the expected improvement in the real data center environments.

\subsection{Traffic Saving}
The parallel transfer by the SDN combined with the non-conventional TCP reception not only saves the time for data write but also the data traffic that crosses the network, which benefits other tasks that share the network resource. We define the total traffic load imposed on the data center network as the volume of the transferred data multiplied by the number of traversed links by it. Given the same data size for the two compared schemes, however, it suffices to count only the number of traversed links $L_{tot}$ for the comparison. In the traditional chain replication, it can be decomposed to

\begin{equation}\label{eq:tot1}
L_{tot} = L_{c,s_1} + L_{s_1,D_1} + L_{D_1,s_2} + \cdots + L_{s_k,D_k}
\end{equation}
If we let $c\equiv D_0$,
\begin{equation}\label{eq:tot2}
L_{tot} = \sum_{j=0}^k \left( L_{D_j,s_{j+1}} + L_{s_{j+1},D_{j+1}}\right)
\end{equation}

Here, $L_{x,y}$ denotes the number of intra-datacenter links from $x$ to $y$, and $\{s_i\}_{i=1}^k$ are the (possibly overlapping) ordered list of switches from which the next descending starts towards the next data node in the pipeline. For example, in Figure \ref{fig:dc-topology}, the list is $\{s_1=s_c,s_2=s_a,s_3=s_c\}$. Note that the first term in the summation is the number of ascending links going up in the hierarchy, whereas the second term is those descending from the switches to the data nodes. In Figure \ref{fig:dc-topology}, for example, the former links are $\{5,7,8,9\}$ and the latter are $\{2,3,4,6,10,11,12\}$ (Link 1 is not in the data center). The idea of the mirrored replication is to eliminate the ascending links in the replication, \textit{i.e.,} the terms $L_{D_j,s_{j+1}}\ (j \ge 1)$. 
Therefore, the traffic saving ratio in the proposed system is essentially the fraction of the ascending links in $L_{tot}$ and the traffic saving ratio is given as

\begin{equation}\label{eq:ratio}
\sum_{j=1}^k L_{D_j,s_{j+1}}/L_{tot}
\end{equation}

\begin{figure}[ht]
\centering
\includegraphics[width=0.95\linewidth]{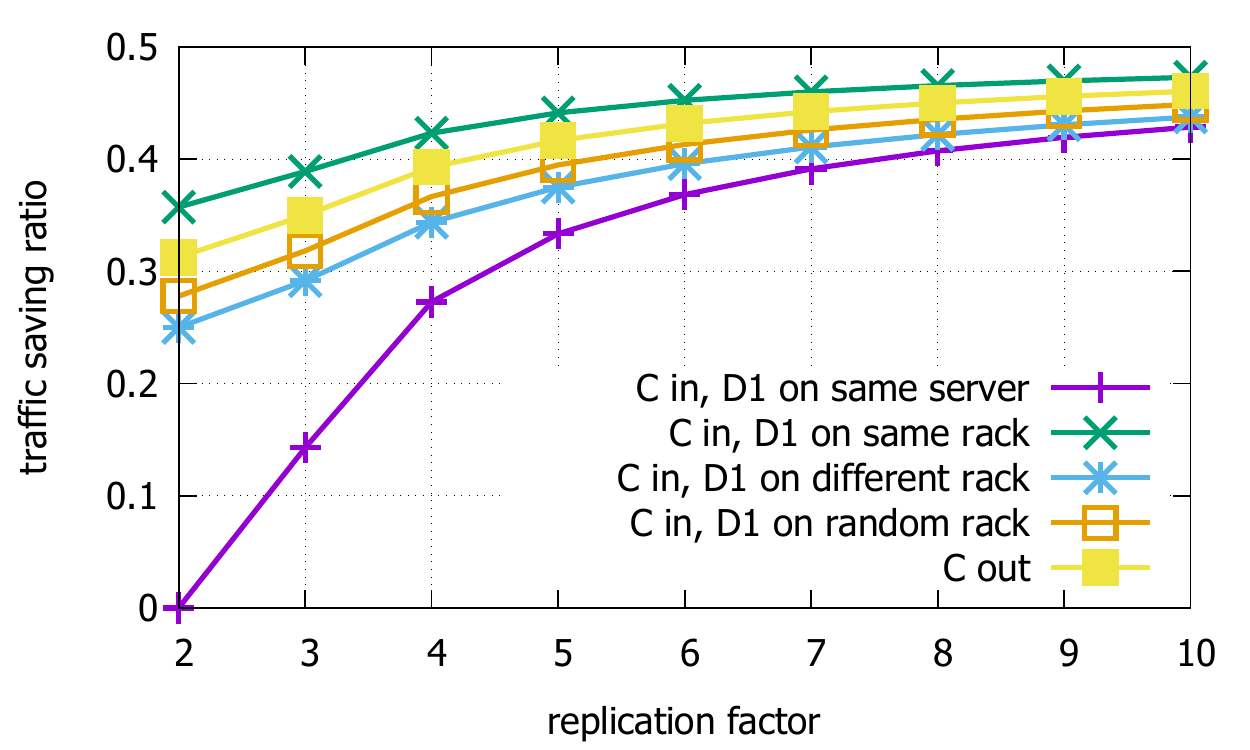}
\caption{Average traffic saving ratios by using the mirrored replication}
\label{fig:traffic-saving}
\end{figure}

In the typical three-layer switching network that comprise edge (top-of-rack), aggregation, and core switches \cite{chen16} as in Figure \ref{fig:dc-topology}, we can consider two cases: the client is either outside or inside the data center network. In the former, $L_{c,s_1}=0$ (we count only the links in the data center network), and the data flow descends through three links. The latter case is further subdivided into four cases. First, if the client colocates with $D_1$ on the same server, $L_{D_0,s_1} = L_{s_1,D_1} = 0$. Moreover, $L_{D_1,s_2}$ cannot be eliminated since it should be the replication source. Second, if the client and $D_1$ are on the same rack but not on the same server, $L_{D_0,s_1} = L_{s_1,D_1} = 1$. Third, if they are located in different racks, $L_{D_0,s_1} = L_{s_1,D_1}$ are either 2 or 3. Otherwise, the number of links can be anywhere between 1 and 3, for both ascending and descending. From $D_4$ and on, there is no constraint, so it is also anywhere between 1 and 3. Given these, Figure \ref{fig:traffic-saving} shows that the average traffic reduction is substantial, ranging from 15 to 40\% at the typical replication factor of 3, and likely more for larger replication factors. 

\section{Discussion}
Our scheme, TCP-MR, is a novel approach in that not only it preserves the pipeline structure during the parallel transfer, as opposed to the previous solutions, in order to reduce data replication time and traffics in the network but also it does stipulate how the SDN switches should work in order to support TCP-MR in the current SDN standard \cite{openflow151}. Although it is known to be beneficial for data centers to adopt a SDN \cite{SDN_BIG_DATA}, its application usually are limited within the network resource management through multicast \cite{ATHENA} \cite{MCTCP}, routing \cite{Avalanche}, switch selection \cite{MCDC}, and so on. Meanwhile, in our scheme, SDN switches modify the TCP/IP header and forward the transparently mirrored TCP segments in order to support the TCP extension in a concrete and unprecedented way.

TCP-MR may be concerned with the absence of TCP flow control between the client and $D_j$ ($2 \le j \le k$), however, this concern is trivial with the sufficient kernel memory for data reception at $D_j$. It prevents $D_j$ from dropping the successfully received out-of-order segments to provide the sufficient kernel memory for data reception with $D_j$. For instance, the mirrored segments from the client may sometimes get lost. Not surprisingly, $D_j$ tries to fill the hole from $D_{j-1}$ by sending the TCP ACK and getting the retransmitted segment. During this recovery process, the mirrored out-of-order segments from the client consumes the receive buffer at $D_j$ quickly and the receiver buffer becomes zero. However, the mirrored out-of-order segments keep coming because of the absence of TCP flow control between them. From this moment on, $D_j$ drops even the successfully received out-of-order segments due to no receive buffer and ask $D_{j-1}$ to retransmit them. With the sufficient receive buffer, however, $D_j$ does not discard any successfully received segments from the client. The sufficient kernel memory size for HDFS is \texttt{writeMaxPackets}$\times$\texttt{HDFS packet size}. \texttt{writeMaxPackets} is the maximum number of HDFS packets to send without the HDFS ACK in the application layer from all data nodes in the pipeline and it is 20 for default. \texttt{HDFS packet} is the unit by which HDFS in the application layer divides the data and its default size is 64 KB. Setting aside the kernel memory, which is set manually for performance evaluation, in TCP-MR is necessary for future work.

It is our future work to develop a congestion-aware routing application, tailored to TCP-MR in SDN to complement the absence of TCP congestion control between the client and $D_j$ ($2 \le j \le k$). In the SDN architecture, network intelligence such as monitoring congested links and routing to avoid those links is easy to implement because the control and data planes are decoupled \cite{SDN_BIG_DATA}. Still, TCP-MR has advantages over conventional TCP because paths (\textit{i.e.,} {10,11,12} in Figure \ref{fig:dc-topology}) taken by TCP-MR is the subset of paths (\textit{i.e.,} {7,8,9,10,11,12} in Figure \ref{fig:dc-topology}) taken by conventional TCP.

Some may argue that TCP-MR is only applicable to HDFS, however, it can be applied to any cluster file systems such as GFS and Windows Azure Storage as long as they exploit sequential TCP connections to replicate data. It may require some minor interfacing for adoption, however, there is no burden to application programmers for cluster file systems once it is adopted. As mentioned the above, we used HDFS as a guiding example.

TCP-MR is not comparable to the scheme establishing TCP subflows such as MPTCP because TCP-MR deals with one to many communication whereas MPTCP deals with one to one communication. TCP-MR could be augmented with MPTCP, however, it is beyond the scope of this paper.

\section{Conclusion}
This paper shows that through a TCP extension, the chain replication used in many cluster file systems 
can be performed in parallel among SDN-connected data nodes. The proposed concept preserves the chain replication semantics, which allows HDFS or Hadoop to work without modification. We show by using a prototype implementation that it improves the data transfer time by 25\% for typical replication factors. Also, it saves 15 to 40\% traffic for the typical replication factor of 3 and more for larger replication factors.

\bibliography{TCP-MR}
\bibliographystyle{IEEEtran}
\end{document}